 \documentclass[aps,prb,twocolumn,footinbib,floatfix]{revtex4}
 \usepackage{graphicx}
\newcommand{\bwt}{\begin{widetext}}
\newcommand{\ewt}{\end{widetext}}
\newcommand{\comentario}[1]{  }
\newcommand{\valorabst}{|t|}
\newcommand{\pupilf}{\, O}

\newcommand{\iint}{\mathop{{\int\!\!\!\!\!\int}}}

\newcommand{\intc}{\mathop{\int\!\!\!\!\!\int\!\!\!\!\!\int\!\!\!\!\!\int}}
\newcommand{\bel}[1]{ \begin{equation}\label{#1}}
\newcommand{\ee}{\end{equation}}
\newcommand{\ev}[1]{E\left( #1 \right)}
\newcommand{\evsq}[1]{E^2\left( #1 \right)}
\newcommand{\Resul}[1]{Result~\ref{#1} of Section~\ref{sec:mathresults}}
\newcommand{\Equa}[1]{Eq.~\ref{#1}}
\newcommand{\figu}[3]{
\begin{figure}
\begin{center}
\includegraphics[width=0.5\textwidth]{#2}
\end{center}
\caption{#3}
\label{#1}
\end{figure}
}

\newtheorem{result}{Result}

\begin{document}
\title{Windowed Defocused Photographic Speckle Vibration Measurement}
\author{Jose Diazdelacruz}
\affiliation{Department of Applied Physics, Faculty for Industrial Engineering, \\ Polytechnic University of Madrid.\\  Jose Gutierrez Abascal 2. 28006 Madrid. Spain}

\begin{abstract}
The out-of-plane vibration of a rough surface causes an in-plane vibration of its speckle pattern when observed with a defocused optical photographic system. If the frequency of the oscillations is high enough, a time-averaged specklegram is recorded from which the amplitude of the vibration can be estimated. The statistical character of speckle distributions along with the pixel sampling and intensity analog-to-digital conversion inherent to electronic cameras degrade the accuracy of the amplitude measurement to an extent which is analyzed and experimentally tested in this paper. The relations limiting the mutually competing metrological features of a defocused speckle system are also deduced mathematically. 
\end{abstract}

\maketitle



\section{Introduction}

This paper introduces and analyzes a technique to measure the amplitude of flexural vibrations of plates and beams by means of time-averaged digital defocused speckle photography. These movements consist of out-of-plane oscillations of the real surface around the mean plane. At each point of the surface there is a tangent plane which is {\em tilted} with respect to the mean one. The tilt angle also oscillates. In a defocused optical system tilts are observed as displacements of the speckle pattern (or specklegram), which after being identified, may determine the tilt angle.    

When the frequency of oscillation is fast enough, the image  of a continuously vibrating surface is the time-average of its speckle pattern. More precisely, it is proportional to the convolution of the still speckle pattern (the one obtained when no oscillation is present) and the inverse of the velocity expressed as a function of the displacement.

Ideally, the statistics of the time-averaged speckle, and, more specifically, of its expected contrast, are determined by the ratio of the vibration amplitude of the recorded pattern to the average speckle size. When a digital camera is used, the integration area of the detector cell reduces the expectation value of the contrast.  Moreover, even if there is a precise relation between  expected contrast and vibration amplitude, the contrast of the sampled specklegram will be a random variable and differ from the population mean; this will make the estimated amplitude deviate from the true one. The fewer speckles in the sampled image, the more variable its contrast will be. Finally, the existence of a finite number of possible gray levels introduces a new source or error.   

In the following sections calibration curves that relate amplitude to contrast and expressions for measurement uncertainty are deduced mathematically. Some results that are necessary to follow the calculations have been displaced to Section \ref{sec:mathresults} in order to improve readability. Then the technique is discussed, the experimental results are presented and some conclusions are drawn.
   
The first paper describing a defocused two-exposure method to measure out-of-plane rotations or tilts was due to Tiziani\cite{tiziani1} and was later extended for vibration analysis\cite{tiziani2}. According to \cite{rastogi}, if normal illumination and observation are used, the speckle shift at the recording plane is given by

\begin{equation}\label{dalf}
d_x= 2 f \alpha
\end{equation}
\begin{equation}\label{dbet}
d_y= - 2 f \beta
\end{equation}
where $\alpha,\beta$ are the (small) rotation angles around the $y,x$ axis of a Cartesian system placed on the mean plane of the object surface and $f$ is the focal length of the recording system. Lateral displacements do not appreciably alter these values.

Gregory\cite{gregory1,gregory2,gregory3} considered divergent illumination and showed that when the optical system is focused on the plane than contains the image of the point source considering the object surface as a mirror, the speckle shift only depends on out-of-plane tilts. Chiang and Juang\cite{chiangjuang} described a method to measure the change in slope by defocused systems. A great number of later papers\cite{platedeformation,fotovibration,schwieger,sjodahl} document the use of defocused speckle photography to measure in-plane and out-of-plane rotations and strains. 

Today CCD cameras store the specklegrams taken before and after the mechanical transformation in a digital computer and adequate algorithms reveal the speckle shift distribution with sub-pixel accuracy\cite{sutton,chen,sjo93,amodio}. 

Electronic Speckle Pattern Interferometry (ESPI) uses the interference of a reference beam and one scattered by a rough surface. It has been used to measure vibration amplitudes\cite{lokberg-espi,shellabear-espi,slangen-espi,wong-espi} and elastic properties of materials\cite{kang} quite accurately, with the limitation that the speckle displacement should not exceed the speckle size. 

The work by Takai\cite{takai} describes the contrast reduction in time-averaged speckle photography of a vibrating surface. His treatment assumes a sinusoidally vibrating speckle pattern and can be applied to imaging and defocused systems as well. However, it does not offer the integrated mathematical form presented on section \ref{generalmaths} in this work and deals with film recorded speckles whereas this paper focuses on digital electronic systems. 

The intensity correlation of time-averaged speckle photographic systems can also be used to measure tilt vibrations in imaging\cite{spagnolo} and defocused \cite{wong-ph} systems. They are based on the  correlation of speckle patterns taken before and during the vibration, appreciated through the subtraction of the recorded specklegrams.

This paper is a natural extension of the tilt determination techniques described in \cite{diazcruz2005,diazcruz2007} to time-averaged vibration measurement. In all three cases the specimen is illuminated by circular beams of radius $a$ and the metrological characteristics of the systems can be adapted to different conditions (i.e. measuring range, resolution, etc.) by setting $a$ and the aperture diameter of the camera to adequate values.

More recently, a new speckle technique for measuring the flow of blood vessels, known as LASCA (Laser Speckle Contrast Analysis), is making extensive use of speckle pattern contrast to determine the fluid velocity\cite{briers-fercher,briers-webster,keene}. 

Other methods are based on the relationship between the geometric moments of blurred images and the movement undergone by the photographed objects \cite{burns-helbig,wang-guan}.


\section{Vibration of the intensity pattern}\label{generalmaths}

\figu{fig:beam1}{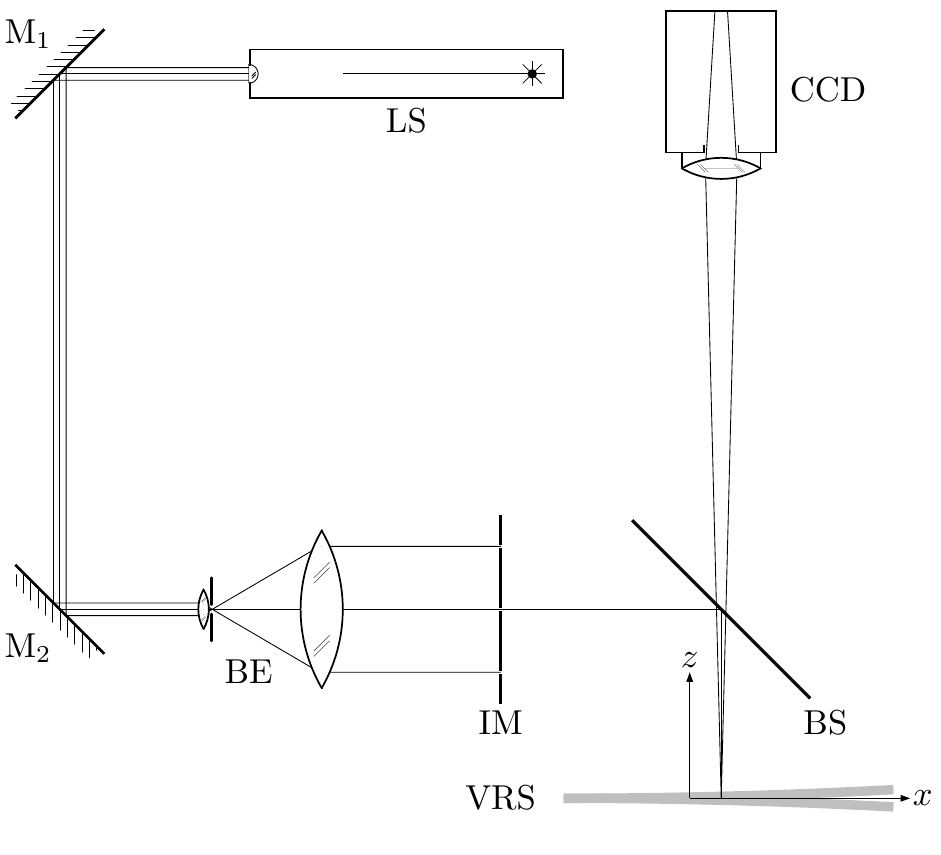}{Optical arrangement: a laser source LS generates a beam that, after being reflected at mirrors M1 and M2, and going through a beam expander BE and a circular window of radius $a$ at the illumination mask IM, is scattered by a vibrating rough surface VRS and recorded by a CCD camera focused at infinity.}

In the optical arrangement depicted in Fig.\ref{fig:beam1} a beam emerging from a laser source LS is reflected at mirrors M1 and M2 and  scattered by a vibrating rough surface VRS after going through beam expander BE and a circular window of radius $a$ at the illumination mask IM. Finally, a CCD camera focused at infinity records the scattered light at its back focal plane, where two cartesian axes $x,y$ are defined. 

If the vibration frequency is $\nu$ and the tilt $\alpha(t)$ is around an axis paralell to $y$, then
\begin{equation}\label{alfadet}
\alpha(t)=\theta \cos 2\pi \nu t
\end{equation}
where $\theta$ is the angular vibration amplitude, which, in general, may be different for each point on the surface. 

At the recording plane the speckle pattern experiences a dynamic displacement of the same frequency and amplitude 
$\zeta$ which, according to \Equa{dalf} and \Equa{dbet}, is related to $\theta$ and the focal length $f$ of the camera through the equation
\begin{equation}\label{relacion-xi-theta}
\zeta = 2 f \theta
\end{equation}
and the velocity of the speckle pattern is along the $x$ axis. It can be expressed by 
\begin{equation}\label{velocity}
\dot \xi = 2\pi \nu \zeta \sin 2\pi \nu t
\end{equation}
so that the probability density for a displacement $(\xi,\eta)$ is
\begin{equation}
\label{proba}
h_2(\xi,\eta) = h_1(\xi)\delta(\eta)
\end{equation}
where
\begin{equation}\label{h1def}
h_1(\xi)=\frac{1}{\pi \, \sqrt{\zeta^2 - \xi^2}}
\end{equation}
for $-\zeta < \xi <\zeta$ and zero otherwise, being $\delta(\eta)$ the one-dimensional Dirac delta function. It is straightforward to verify the normalization equation
\begin{equation}\label{hintegral}
\iint h_2(\xi,\eta)\, d \xi\, d \eta =1
\end{equation}

The resulting time-averaged intensity distribution is the convolution of the still pattern $i_0(x,y)$ and $h_2(x,y)$
\begin{equation}
\label{intensity}
i(x,y)=i_0(x,y) \otimes h_2(x,y)  
\end{equation}
\footnote{Throughout this paper $\otimes$ represents convolution with respect to the independent variables shared by its factors, i.e., $F(x,y)\otimes G(x,y)$ is understood as the two-dimensional convolution of $F(x,y),G(x,y)$ with respect to $x,y$, whereas $H(x)\otimes G(x,y)$ is the one-dimensional convolution of $H(x),G(x,y)$ with respect to $x$. }As a consequence of the vibration, the speckle pattern is blurred. The contrast decrease can be accounted for after considering some statistical properties of the intensity distribution, which is assumed to be a spatially stationary two-dimensional stochastic process obeying second order statistics as described in the pioneering work by Goodman \cite{goodman}.  In short, the speckle is a Wide Sense Stationary process with an autocorrelation function:
\begin{equation}\label{fromGoodman}
R_0(\xi,\eta)\equiv E\left(i_0(x+\xi,y+\eta)i_0(x,y)\right)
=I^2 \left(
1+\left| \mu(\xi,\eta) \right|^2
\right)
\end{equation}
where $I$ is the average intensity of the still speckle pattern, $E(F)$ henceforth represents  the statistical expectation value of function $F$ and $\mu(\xi,\eta)$ is the amplitude autocorrelation which, according to \cite{goodman}, is given by
\begin{equation}\label{aucs}
\mu(\xi,\eta)= \lambda f \frac{J_1\left(\displaystyle\frac{2\pi a\displaystyle\sqrt{\xi^2 + \eta^2}}{\lambda f}\right)}{\pi a \sqrt{\xi^2+\eta^2}}
\end{equation}
where $J_1$ is the Bessel function of first kind and order 1, $\lambda$ is the laser wavelength and $f$ is the focal length of the camera\footnote{Actually, the value of $a$ in \Equa{aucs} is the radius of the limiting aperture in the system, that in this paper is assumed to be $a$ (i.e. $2a$ will always be smaller than the diameter of the camera pupil).}.

The speckle size $s$ is usually defined as the diameter of the central peak of the autocorrelation function. If $j_{1,1}$ is the first zero of the $J_1$ Bessel function  ($j_{1,1}\approx 3.8317 $), then 
\begin{equation}\label{az1}
s=\frac{j_{1,1} \lambda f}{\pi a}\approx \frac{1.2 \lambda f}{a}
\end{equation}

The autocorrelation of $i(x,y)$ is
\begin{equation}\label{au}
R(\xi,\eta)=E\left(
i(x+\xi,y+\eta)i(x,y)
\right)
\end{equation}
which, according to \Resul{r2} (applied twice because of the two-dimensional convolution and considering that $h_2(\xi,\eta)=h_2(-\xi,-\eta)$) can be written as
\begin{equation}\label{dobleconv}
R(\xi,\eta)=h_2(\xi,\eta)\otimes h_2(\xi,\eta) \otimes R_0(\xi,\eta)
\end{equation}
but, substituting from \Equa{proba}
\begin{equation}\label{inter11}
h_2(\xi,\eta)\otimes h_2(\xi,\eta) = \left[ h_1(\xi) \otimes h_1(\xi)\right] \delta(\eta)
\end{equation}
so that
\begin{equation}\label{dobleconvr}
R(\xi,\eta)=h_1(\xi)\otimes h_1(\xi) \otimes R_0(\xi,\eta)
\end{equation}
Let a new function $H_1$ be defined by $H_1(\xi) = h_1(\xi) \otimes h_1(\xi)$. Applying  \Resul{ellipticK}, one can write, for $-2\zeta \leq \xi \leq 2\zeta$,
\begin{equation}\label{red}
H_1(\xi) =  \frac{4}{\pi^2 \left(
2\zeta + \left|
\xi
\right|
\right)}
K\left(\displaystyle{
\frac{2 \zeta - |\xi|}
{2 \zeta + |\xi|}
}\right)
\end{equation}
and 0 otherwise, where $K$ is the complete elliptic integral of the first kind.
Accordingly, 
\begin{equation}\label{au4p}
R(\xi,\eta)=
H_1(\xi) \otimes R_0(\xi,\eta)   
\end{equation}

The autocovariance of the vibrating speckle pattern is
\begin{equation}\label{defC}
C(\xi,\eta)=R(\xi,\eta)-I^2
\end{equation}
which, according to \Equa{fromGoodman} and \Equa{au4p} yields
\begin{equation}\label{autocorrelation}
C(\xi,\eta)= \iint
I^2 \left( \left|\mu(\xi+u,\eta)\right|^2 + 1\right) H_1(u)  du - I^2
\end{equation}
and using the probability normalization of \Equa{hintegral} can be written as
\begin{equation}\label{CC}
C(\xi,\eta)= \int
I^2 \left|\mu(\xi+u,\eta)\right|^2 H_1(u) du
\end{equation}
If a contrast index $Q$ is defined by 
\begin{equation}\label{Q}
Q=\sqrt \frac {C(0,0)}{I^2}
\end{equation}
then 
\begin{equation}\label{Q2}
Q^2 = \int
\left|\mu(u,0)\right|^2 H_1(u) du 
\end{equation}
is only dependent on the speckle amplitude autocorrelation and the maximum vibration displacement. Substituting from \Equa{aucs} and \Equa{red}, after the change of variable $\xi = 2\zeta w$,
one obtains
\begin{equation}\label{tochon}
Q^2 =\frac{8}{\pi^2} \int_{0}^{1} \frac{1}{\left(
1 + 
w
\right)}
K\left(\displaystyle{
\frac{1 - w}
{1 + w}
}\right) \left(\frac{J_1\left(\displaystyle\frac{4 j_{1,1}\zeta w}{s}\right)}{\displaystyle\frac{2 j_{1,1}\zeta w}{s}}\right)^2 d w
\end{equation}
that only depends on the adimensional parameter $\zeta/s$.

\figu{fig:graf}{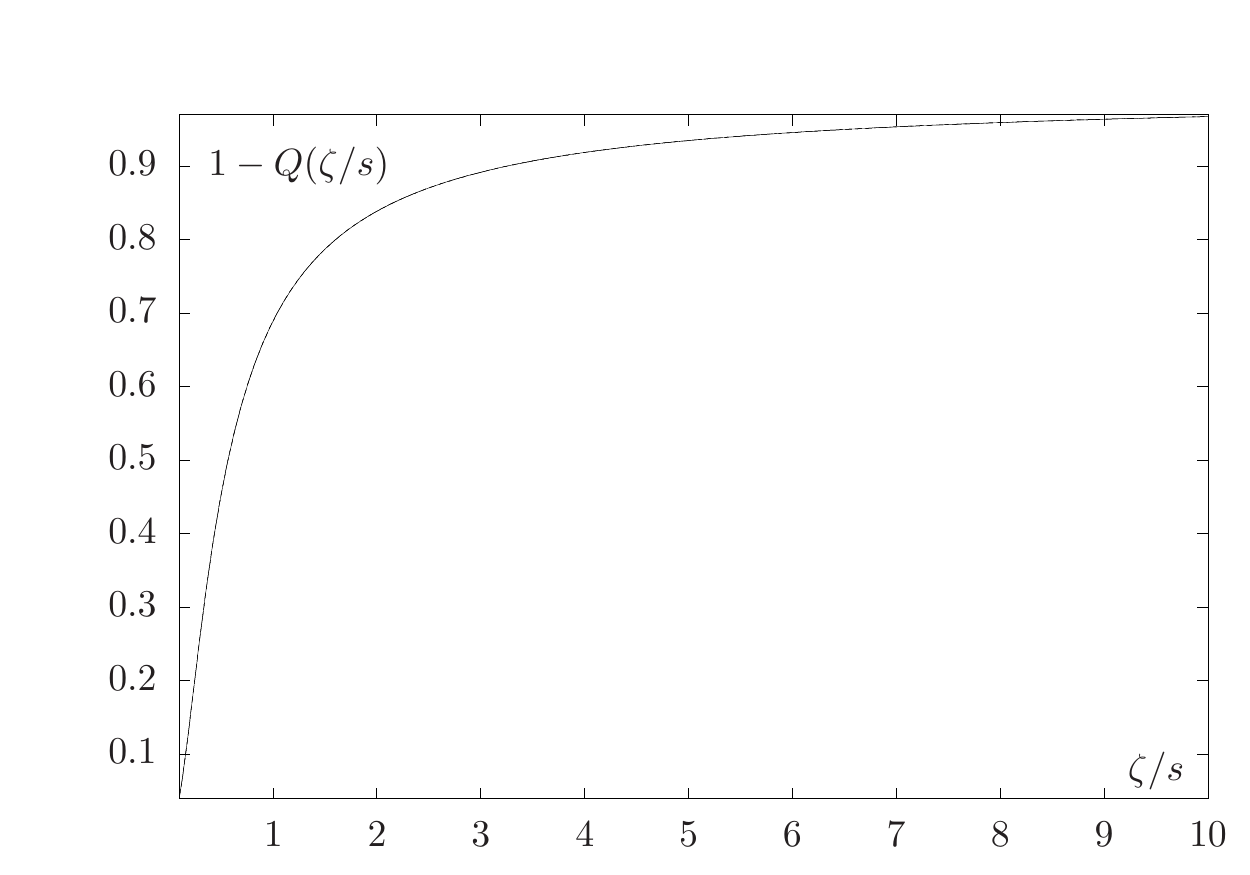}{Graphical representation for $1-Q(\zeta/s)$ as a function of the adimensional parameter ${\zeta}/{s}$ (vibration amplitude on the recording plane over speckle size)}

The function $1-Q(\zeta/s)$ is represented by a solid line in Fig.\ref{fig:graf} which reveals that $Q$ approaches 1 if the speckle is still. As the vibration amplitude increases, the speckle is blurred and its contrast decreases. When the amplitude is much greater than the speckle size the image approaches an uniform gray pattern. 

\comentario{
The sensitivity of $Q$ towards the vibration amplitude is small for short vibrations where $Q\approx 1$ or large ones when $Q\approx 0$.
}


\section{Incremental contrast}

In the previous section a contrast index $Q$ has been defined for a time-averaged speckle pattern $i(\xi,\eta)$ recorded on a plane $\xi,\eta$. It has been related to the adimensional parameter $\zeta/s$ by means of \Equa{tochon}. However, tracing back the definition of the index $Q$ through \Equa{Q},\ref{defC},\ref{au}, it can be written as
\begin{equation}\label{tb}
Q^2 = \frac{\ev{i^2(x,y)} - \evsq{i(x,y)}}{\evsq{i(x,y)}}
\end{equation}
from which it is evident that $Q$ has been defined using the expectation values of $i^2(x,y), i(x,y)$. If a finite speckle pattern is sampled, $Q^2$ has to be estimated from the recorded data and the process will introduce some uncertainty which will be quantified in this section. Using an aperture of diameter $D$, a focal length $f$ on the optical system, and a vibrating surface placed at a distance $d$ from the objective of the camera, the speckle pattern from an illuminated spot will be on a circle of radius $c$, given by
\begin{equation}\label{diameter}
c = \frac{D f}{2d}
\end{equation} 

It is evident that the estimation of $Q^2$ from the distribution of $i(\xi,\eta)$ on a circle of 
radius $c$ contains a high uncertainty unless $s \ll c$.   A less random  parameter is the {\em incremental contrast} defined by
\begin{equation}\label{defP}
P=\displaystyle \frac{1}{\pi c^2 I^2}\left(
\iint_c i_0^2 d S - \iint_c i^2 d S
\right)
\end{equation}
whose variability will be considered later in this section. The integrals extend to a circle of radius $c$. 
The expectation value for $P$ is
\begin{equation}\label{P1}
\ev{P} = Q^2(0)-Q^2\left(\frac{\zeta}{s}\right)
\end{equation}
which, according to \Equa{tochon} can be written as
\begin{equation}\label{tochonon}
\ev{P} = 1- \frac{8}{\pi^2} \int_{0}^{1} \frac{1}{\left(
1 + 
w
\right)}
K\left(\displaystyle{
\frac{1 - w}
{1 + w}
}\right) \left(\frac{J_1\left(\displaystyle\frac{4 j_{1,1}\zeta t}{s}\right)}{\displaystyle\frac{2 j_{1,1}\zeta w}{s}}\right)^2 d w  
\end{equation}

As stated earlier, different sampled circles may determine different evaluations for $P$, specially if the radius $c$ is not much larger than the speckle size. Next, in order to approach the variance of $P$, a linearization of $i^2$ around $I$ is made
\begin{equation}\label{az5}
i^2 \approx 2 I i - I^2
\end{equation}
so that the linearized incremental contrast
\begin{equation}\label{az6}
P'=\frac 2 {\pi c^2 I} \iint_c (i_0-i) d x d y
\end{equation}
is considered, whose variance is
\bwt
\begin{equation}\label{varpprime}
VAR(P') = \frac {4}{\pi^2 c^4 I^2}  \iint_c \iint_c \ev{\left[i(x,y)-i_0(x,y)\right]\left[i(x',y')-i_0(x',y')\right]} d x d y \,\, d x' d y' 
\end{equation}
\ewt

In order to use the \Resul{r2} to evaluate \Equa{varpprime} it must be taken into account that $ i(x,y) - i_0(x,y) =  \left[ h_2(x,y)-\delta(x,y)\right] \otimes i_0(x,y)$. It follows that
\bwt\begin{equation}\label{az7}
VAR(P') = \frac{4}{\pi^2 c^4}  \iint_c\iint_c \left\{\left[h_1(\xi) \otimes h_1(\xi) - 2 h_1(\xi) + \delta(\xi)\right] \delta(\eta) \right\}\otimes \left|  \mu(\xi,\eta)\right|^2 d x d y \,\, d x' d y'  
\end{equation}\ewt
being $\xi=x-x'$, $\eta=y-y'$, or
\begin{equation}\label{az8}
VAR (P') = \frac{4}{\pi^2 c^4}  \iint_c\iint_c \left[ G(\xi) \otimes (| \mu(\xi,\eta)|^2\right]  d x d y \,\, d x' d y' 
\end{equation}
where
\begin{equation}\label{az9}
G(\xi)=H_1(\xi)-2 h_1(\xi) +\delta(\xi)
\end{equation}
Considering that $G(\xi) \otimes (| \mu(\xi,\eta)|^2$ only depends on $(x-x',y-y')$, \Resul{r1} is applicable and yields

\begin{equation}\label{varPp}
VAR (P') = \frac{4}{\pi c^2}  \iint \left[ G(\xi) \otimes | \mu(\xi,\eta)|^2\right] L(\xi,\eta) d \xi  d \eta 
\end{equation}
\figu{fig:unc}{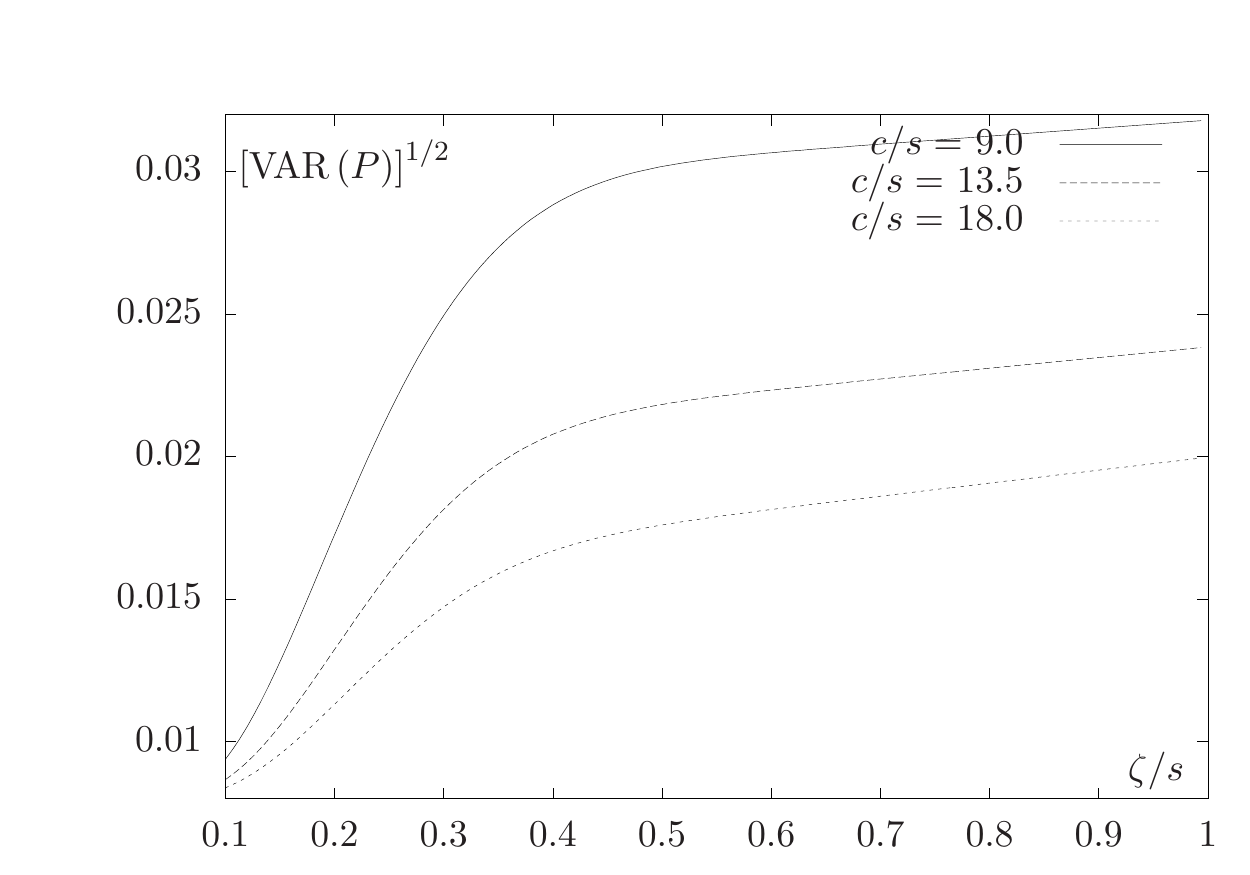}{Graphical representation for the stardard deviation of the incremental contrast $P$ sampled for different $c/s$ ratios as a function of $\zeta/s$}
Fig.\ref{fig:unc} represents the stardard deviation (the square root of $VAR(P')$) for different $c/s$ ratios as a function of $\zeta/s$.

\comentario{
In order to have an estimation of the uncertainty that the variance of $P'$ causes in the determination of the vibration amplitude $\zeta$, the derivative of $P(\zeta)$ is written
\begin{equation}\label{DP}
P_\zeta \approx \frac{\Delta P}{\Delta \zeta}
\end{equation}
whence
\begin{equation}\label{DPa}
VAR(\zeta) \approx \frac{VAR(P)}{P_\zeta^2} \approx  \frac{VAR(P')}{P_\zeta^2}
\end{equation}

}


\section{Digital recording of the blurred speckle}

The extended character of the sensor cells in the recording camera introduces an averaging action on the speckle which may further reduce its contrast. This effect can be accounted for by considering the function $i'(x,y)$ that assigns to every point on the recording plane the average of the incident intensity $i(x,y)$
on the sensor area, which is considered to be a square of side $b$. Accordingly, one can write 
\bwt\begin{equation}
\label{iprime}
i'(x,y) = i(x,y) \otimes \left[\frac 1 {b^2} \Pi\left(\frac x b \right) \Pi\left(\frac y b \right)\right]= i_0(x,y) \otimes h_1(x)  \delta(y)  \otimes \left[\frac 1 {b^2} \Pi\left(\frac x b \right) \Pi\left(\frac y b \right)\right]
\end{equation}\ewt
being $\Pi(x)$ the {\em rectangular function} given by $\Pi(x) = 1$ if $2|x|\leq 1 $ and 0 otherwise. 

Let $R'(x,y)$ be the autocorrelation function of $i'$. The expected value of $i'^2(x,y)$ is 
\begin{equation}\label{evip}
\ev{i'^2}=\ev{i'(x,y)i'(x,y)}=R'(0,0)
\end{equation}
where, referring to \Resul{r2}
\bwt\begin{equation}\label{evip1}
R'(x,y)= \frac{1}{b^4} R_0(x,y)\otimes \left\{\left[ (h_1(x) \otimes h_1(x) \otimes \Pi\left(\frac x b \right)\otimes \Pi\left(\frac x b \right)\right]\left[
 \Pi\left(\frac y b \right) \otimes  \Pi\left(\frac y b \right)
\right]
\right\}
\end{equation}\ewt
and applying \Resul{ellipticK}
\bwt\begin{equation}\label{evip2}
R'(x,y)= \frac{1}{b^4} R_0(x,y)\otimes \left\{\left[ H_1(x)\otimes \Pi\left(\frac x b \right)\otimes \Pi\left(\frac x b \right)\right]\left[
 \Pi\left(\frac y b \right) \otimes  \Pi\left(\frac y b \right)
\right]
\right\}
\end{equation}\ewt
Let $\Lambda(x)$ be the {\em triangular function} given by
\begin{equation}\label{Ldef}
\Lambda(x)=\left\{
\begin{array}{l}
1-|x| \mbox{ if } |x|\leq 1 \\
0 \mbox{, otherwise}
\end{array}
\right.
\end{equation}
Then it follows that 
\begin{equation}\label{pipilamb}
\frac 1 {b^2} \Pi\left(\frac x b \right)\otimes \Pi\left(\frac x b \right) = \frac{1}{b}\Lambda(\frac{x}{b})
\end{equation}
and thus
\begin{equation}\label{evip11}
R'(x,y)= \frac{1}{b^2} R_0(x,y)\otimes \left\{
\Lambda\left(\frac{x}{b}\right)\Lambda\left(\frac{y}{b}\right)
\right\}
\otimes H_1(x)
\end{equation}
which leads to
\begin{equation}\label{evip11e}
\ev{i'^2}= \frac{1}{b^2} \iint \left[H_1(x)\otimes \Lambda\left(\frac{x}{b}\right)
\right]  R_0(x,y) \Lambda\left(\frac{y}{b}\right) d x d y 
\end{equation}
so that 
\begin{equation}\label{evip11ee}
{Q^2}= \frac{1}{b^2} \iint \left[H_1(x) \otimes \Lambda\left(\frac{x}{b}\right)
\right]|\mu(x,y)|^2  \Lambda\left(\frac{y}{b}\right) d x d y 
\end{equation}
finally, taking into account that
\begin{equation}\label{P1bis}
\ev{P} = Q^2(0)-Q^2\left(\frac{\zeta}{s}\right)
\end{equation}
\Equa{evip11ee} yields
\begin{equation}\label{evip11eeu}
\ev{P}=  \frac{1}{b^2} \iint \left[\left\{\delta(x)-H_1(x)\right\} \otimes \Lambda\left(\frac{x}{b}\right)
\right]  |\mu(x,y)|^2 \Lambda \left(\frac{y}{b}\right) d x d y 
\end{equation}
that depends on $\zeta/s,b/s$. Fig.\ref{fig:calib} represents $\ev{P}$ as function of $\zeta/s$ for different $b/s$ ratios.

\figu{fig:calib}{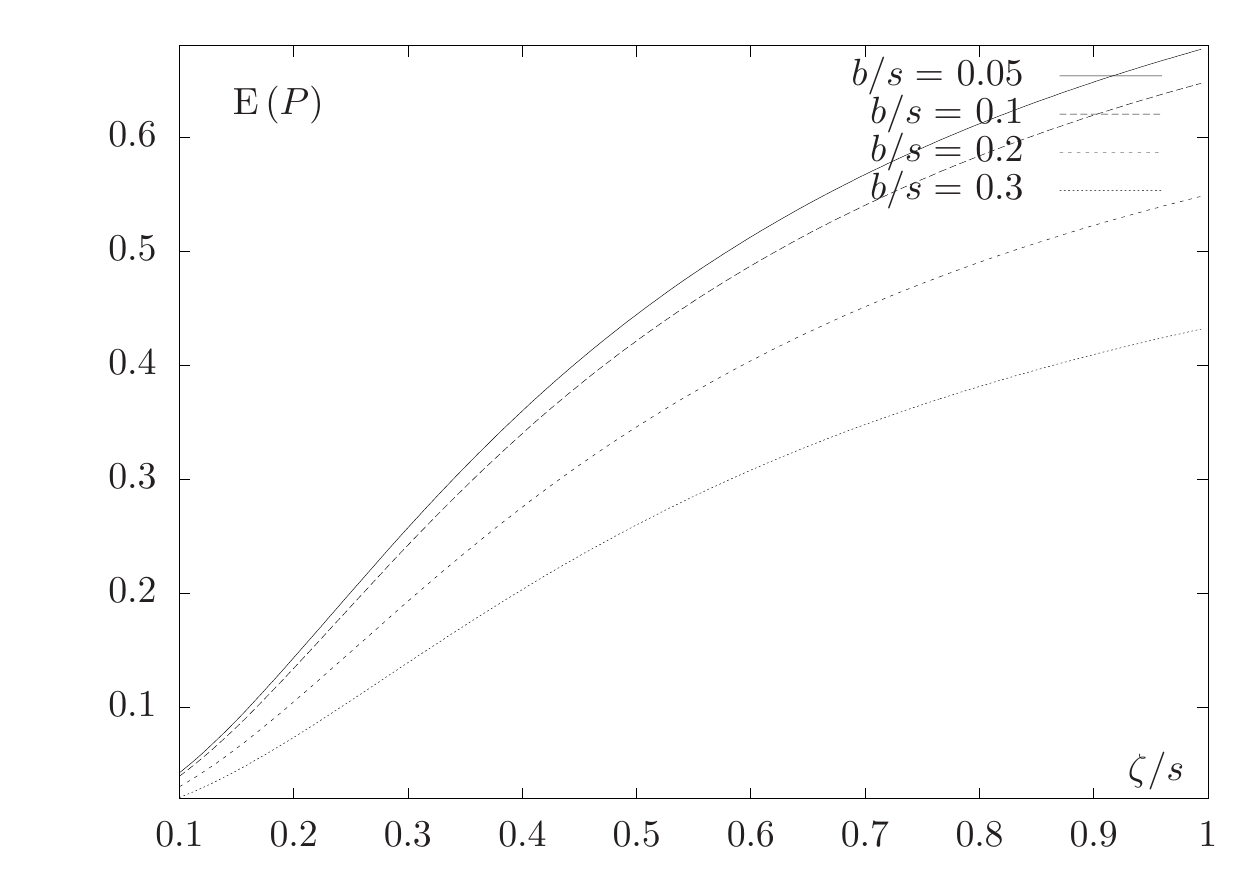}{Graphical representation for the expected value of the incremental contrast $P$ as a function of $\zeta/s$ for different values of $b / s$}
There is another source of information degradation when using monochrome digital cameras. If there are $G$ levels of gray and $N$ sensor cells, a relative error in the order of
$
G^{-1}
$
can be assumed for the intensity $i_0(x,y)$. If the expected contrast is $Q<1$, then the {\em effective} gray levels $G'$ are reduced. It will be assumed that 
$
G'=G Q
$
so that the relative uncertainty in the measured values of $i(x,y)$ is 
$
(GQ)^{-1}
$
and in $i^2$ is 
$
2(GQ)^{-1}
$, which will be taken as a multiplicative factor in the uncertainty of $P$ too (this is a conservative assumption that will be observed in this paper).


\section{Discussion}

This section is focused on the analysis of the metrological properties of the incremental contrast for the determination of the vibration amplitude in rough surfaces.

According to \Equa{evip11eeu}, the expected incremental contrast of the speckle pattern depends mainly on the ratios of two lengths to the speckle size $s$:
\begin{itemize}
\item[1] the normalized vibration amplitude $\zeta$. When the surface under investigation is vibrating the time averaged speckle pattern is blurred. The incremental contrast grows with $\zeta / s$. Yet, if $\zeta / s$ is too low or too high the quality of the vibration amplitude determination is poor. When $\zeta / s$ is small, there is little variation of the contrast and a high uncertainty in the measurement. When $\zeta / s$ is big, the contrast falls to zero (this follows naturally from the averaging of a large number of speckles) and the uncertainty is also is also high. The effect of the amplitude on the incremental contrast is only appreciable and resolvable in the interval  $0.1 \leq \zeta / s \leq 1$. After substitution from \Equa{relacion-xi-theta}, the {\em measuring range} for the angular amplitude $\theta$ can be written as
\begin{equation}\label{mr}
\frac{s }{20 f} \leq \theta \leq \frac{s}{2 f} 
\end{equation} 


\item[2] the width $b$ of the sensor cell. The output of the sensor cell is the average of the intensity in a $b\times b$ square.  The bigger $b / s$, the smaller the differences from the average intensity. The speckle pattern keeps a significant contrast if $ b / s \leq 0.5$; that is, one speckle must cover at least two pixels in each dimension.
\end{itemize}

Following \Equa{varPp}, the uncertainty of the differential contrast depends mainly on:

\begin{itemize}

\item[1] the radius $c$ of the sampled circle. It is clear that the smaller the ratio $ c / s$ (i.e. the less speckles in the sampled circle), the more randomness there is in the statistics of the recorded speckle pattern, specially if $\zeta / s$ is big, because the speckles unrecorded in the still pattern enter the vibrating one. In the experimental part of this work $ c / s$ was never below 9. 

\item[2] the amplitude of the vibration $\zeta$. As it has been stated in the previous item, the vibrating speckle pattern includes the contribution from areas not recorded in the still one, so that large values of $\zeta$ determine high uncertainties.

\item[3] the contrast $Q$. The digital character of the measurement makes it sensitive to the separation of the levels of intensity of the pattern. In a low contrast image the sensor may not resolve for the relatively small differences in intensity. For low contrast images the uncertainty is approximately inversely proportional to the product $ Q G$ where $G$ is the number of levels of gray of the sensor.

\end{itemize}

From the analysis above and the graphical representations in Fig.\ref{fig:unc} it is clear that the speckle size $s$ plays the most important role in configuring the system. It results from \Equa{az1} from which it follows that it can be varied through the aperture $a$ or the focal length $f$.

\figu{fig:relunc}{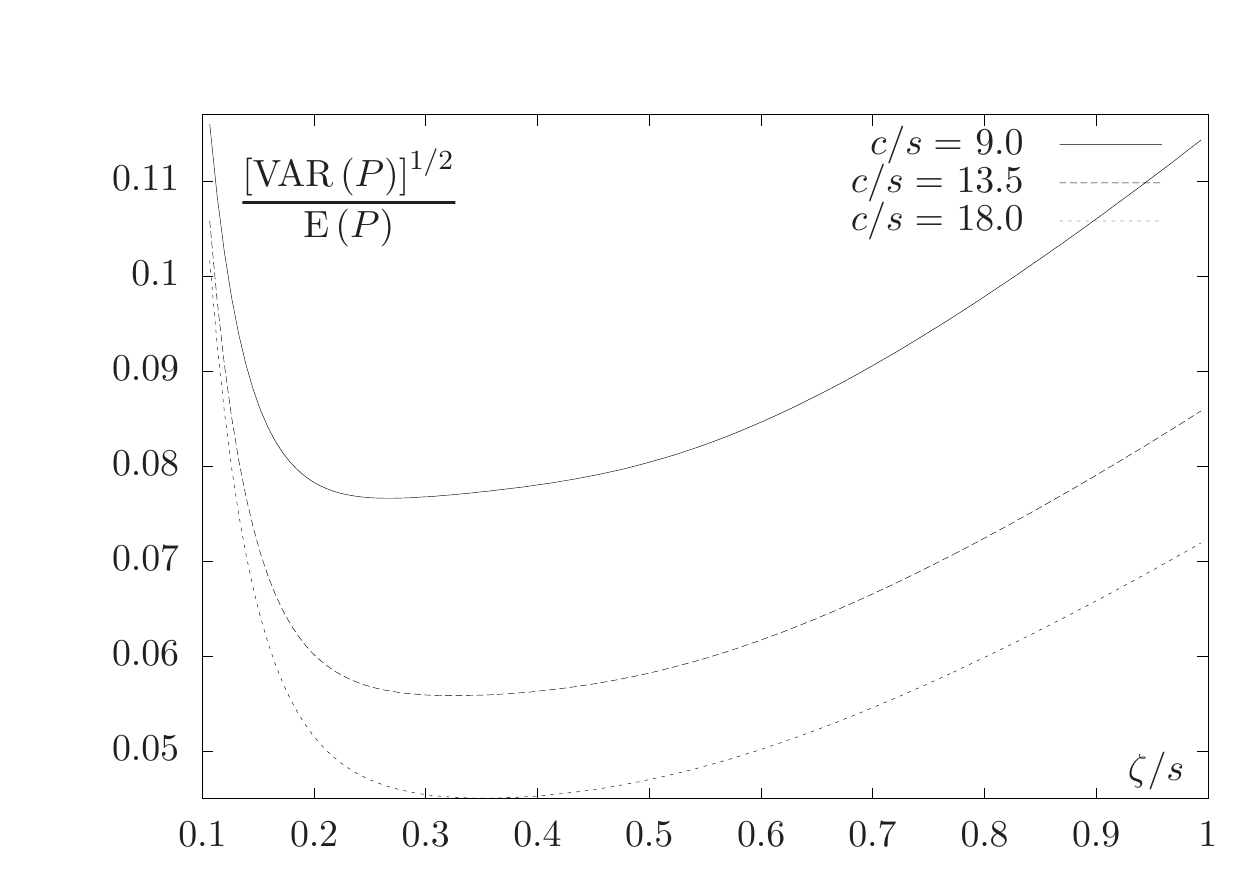}{Graphical representation for the relative uncertainty (standard deviation over expected value) of $P$ as a function of $\zeta/s$ for different values of $c/s$ and $b/s=0.095$}

Fig.\ref{fig:relunc} represents the relative uncertainty in the determination of $P$. There is a central plate where the determination of $\zeta$ can be made with an accuracy that depends mainly on $c/s$. When $\zeta \gg s$ the speckle image is so blurred that it is practically useless for the measurement of $\zeta$. On the other hand, when $\zeta \ll s$, the contrast is not sensitive enough to the vibration amplitude to provide a good determination of $\zeta$.

By reducing the aperture $D$, the radius $c$ of the speckled circle originating from an illuminated spot at the surface also decreases (see \Equa{diameter}) leaving more space on the sensing area of the CCD for the observation of the vibration of more points on the surface. If $\Delta$ is the distance between two illuminated points on the vibrating surface, the separation $e$ of the centers of their speckle circles should be bigger than $2c$, so that  
\begin{equation}\label{separa}
e=\frac{f \Delta}{d} \geq 2c = \frac{D f }{d}
\end{equation}
Therefore,  the minimum distance between two illuminated points so that their speckle circles do not overlap is $D$ and represents the {\em horizontal resolution}.  From \Equa{diameter} it is clear that by reducing $D$ the uncertainty of the measurement is increased. Accordingly, there must be a compromise between the accuracy and the horizontal resolution of the system. The solution will set a value for $D$ so that the speckled circles coming from different illuminated spots do not overlap. This possibility of improving a measuring feature at the cost of degrading other is an advantageous characteristic of the method analyzed on this paper.

Finally, the maximum measurable angular amplitude determines the radius $a$ of the window in the illumination mask. From \Equa{mr}, it is clear that the larger the length of $a$, the smaller the speckle size and the less uncertainty (provided that $s>2b$), but the narrower the measuring range.


\section{Experimental results}

In order to test the foregoing equations for measuring the vibration amplitude of a rough surface, a well-controlled oscillation experiment was set up. A thin mirror plate specular on one side and rough on the other was stuck to a solid pendulum near the fulcrum. 

The period of the small oscillations of the pendulum (with the mirror attached) was $T=1.7$\,s and its relaxation time (interval in which the amplitude oscillation halves) was $\tau\approx 579$\,s. The overall length of the pendulum was $L=1.19$\,m (in this case the usual relation between the period and the length did not hold, because it was a solid instead of a simple pendulum). The rotation angle $\alpha(t)$ of the pendulum followed \Equa{alfadet} with $\nu=1/T$ and the measurement of its amplitude $\theta$ (or the directly related $\zeta=2f\theta$) for a rough surface is what the whole system was set out to accomplish.

A red He-Ne laser source (632.8 nm wavelength) was used and a 8-bit monochrome CCD camera with  752 $\times$ 582 effective output pixels, 11.6 $\times $ 11.2 $\mu$m cell size, and 8.8 $\times$ 6.6 mm sensing area was connected through an image capture board to a 2 GByte RAM and 1.66GHz personal computer.

Before starting the oscillations a specklegram of the illuminated circle on the rough side was taken and the speckle size $s$ measured. In every trial, the bob of the pendulum was slightly drawn aside from its rest position. After releasing it, a variable amount of time was waited until the oscillation approximately reached a target value. Then,  time-averaged specklegrams,  corresponding to a 3.4\,s exposure, were taken for different amplitudes $\zeta$. From each of them three different circles of radius $9 \times s,13.5\times  s, 18 \times  s$ were selected and their contrast $Q( \zeta / s)$ evaluated. \comentario{The aperture of the camera was always maximally open in order to gather the maximum information for later processing.} The results were labeled with the three adimensional parameters $\zeta/s,c/s,b/s$ ($b$ was taken as 11.4$\mu$m). In each test the value of the incremental contrast $P$ was obtained and then, using the corresponding calibration curve in Fig.\ref{fig:calib}, an estimated value of $\zeta / s$ was read. 

The specular side of the mirror was used to provide precise reference values for the oscillation amplitudes. This was accomplished by a second camera that recorded the images of a set of fixed points reflected in the specular side. When the mirror is rotated through an angle $\alpha$ about the fulcrum, the reflected image of a point is rotated through $2\alpha$ about the same axis. When $\alpha$ is small, the recorded image is displaced by a distance approximately proportional to $\alpha$. The corresponding coefficient $k$ was experimentally calibrated. When the oscillation amplitude is $\theta$, the image of a reference point moves along a segment of length $2\, k \, \theta$, from which, taking into account \Equa{relacion-xi-theta}, the reference value for $\zeta$ was obtained.

Experimental results have been represented for two $b/s$ values, obtained by changing the illumination aperture $a$ and the focal length $f$ that determined the speckle size $s$ according to \Equa{az1}. For each, three different oscillation amplitudes and three ratios $c/s$ were tested. 

In order to account for other factors (ambient or stray light, residual vibrations, non-linearity of the detector, etc.), after the images were recorded, the minimum level of gray was subtracted from every pixel and the measured contrast was multiplied by a correction factor  calculated so that the contrast of the still image would match the one from \Equa{evip11ee}.

\figu{fig:b1}{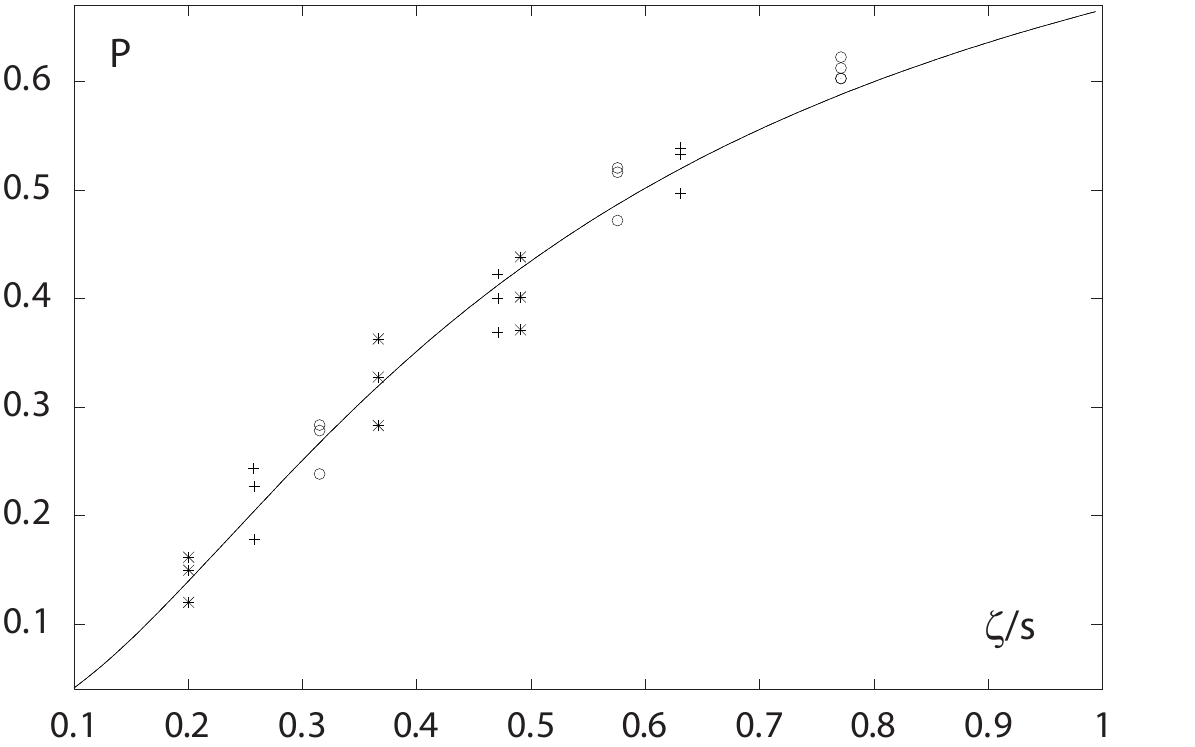}{Results for $b/s=0.095$. The expected value of the incremental contrast is  plotted as a solid line.  The stars, crosses and circles represent the measured values for $c/s = 9,13.5,18$, respectively.  For each $c/s$ ratio, the incremental contrast was determined on three independent experiments for  three $\zeta/s$ values}

Figures \ref{fig:b1} and \ref{fig:b2} plot the experimental results for $b/s=0.095,0.31$, respectively. In each one, the calibration curve from \Equa{evip11eeu}, that represents the incremental contrast as a function of the oscillation amplitude, is plotted as a solid line, whereas the stars, crosses and circles represent the measured incremental contrast for  $c/s=9, 13.5, 18$, respectively. For each value of $c/s$, three different amplitudes were used and for each amplitude three independent measurements were performed. Both figures show how the measured values are distributed around the calibration curves
.
\figu{fig:b2}{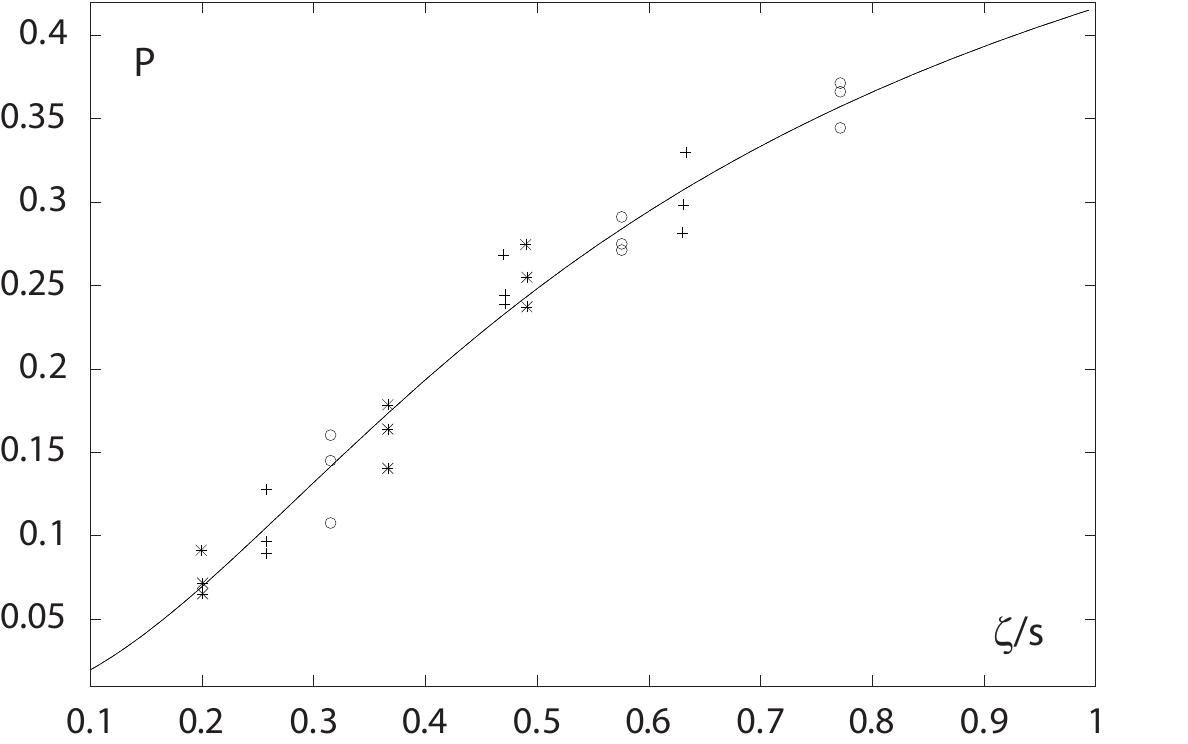}{Results for $b/s=0.31$. The expected value of the incremental contrast is  plotted as a solid line.  The stars, crosses and circles represent the measured values for $c/s = 9,13.5,18$, respectively.  For each $c/s$ ratio, the incremental contrast was determined on three independent experiments for  three three $\zeta/s$ values}

Figure \ref{fig:e1} shows the curves of the standard deviation, as deduced theoretically (\Equa{varPp}), for the three different $c/s$ ratios (solid line, dash-dotted and dashed for $c/s=9.0,13.5,18.0$, respectively)  and the experimental errors in the data obtained  for $b/s=0.095$ (stars, crosses and circles for $c/s=9.0,13.5,18.0$, respectively). The standard deviation in a measuring method is often used to provide a value for the uncertainty of the measurement (sometimes multiplied by 2 or 3). The purpose of the derivation of \Equa{varPp} was precisely to estimate the uncertainty of the incremental contrast determination from a single specklegram. Fig.\ref{fig:e1} shows that the errors are all within three times the standard deviation."

The experimental outcomes show agreement not only with the calibration curve, but also with the limits represented by the computed uncertainty, which result remarkably fitted to the technique described.

\figu{fig:e1}{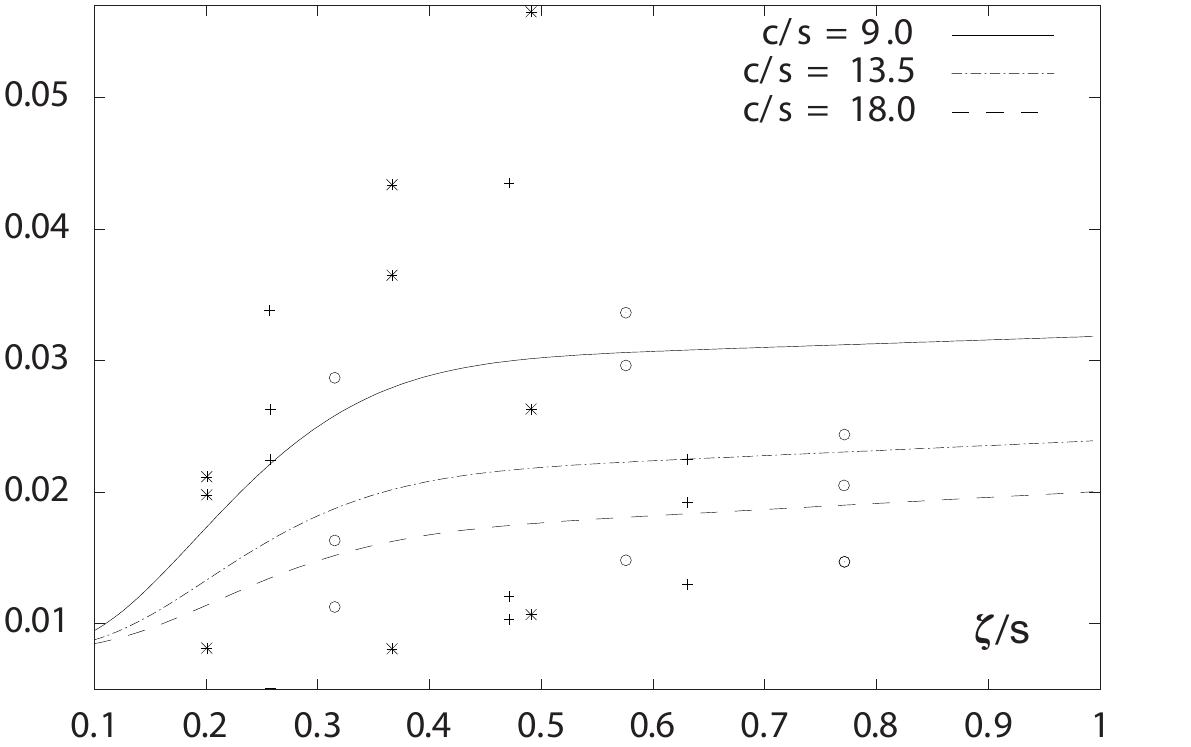}{Deviations for $b/s=0.095$. The  theoretical standard deviation of the incremental contrast is plotted as a function of $\zeta/s$ for the three different $c/s$ ratios (solid line, dash-dotted and dashed for $c/s=9.0,13.5,18.0$, respectively) . The experimental errors appear as stars ($c/s=9$), crosses ($c/s=13.5$) and circles ($c/s=18$)}





\section{Conclusions}

Speckle contrast is significantly sensitive to vibration amplitude in photographic defocused systems only over a narrow interval around the speckle size. In an optical set-up as the one described here, the speckle size can be varied by altering either the focal length or the radius of the circular window in the Illumination Mask. 

The speckle contrast referred to in the previous paragraph is not the one of a recorded finite size specklegram, but the one expected in an ideally infinite recording area. An image with a small number of speckles is likely to have an actually measured contrast significantly different from its expected value. This source of error can be reduced by increasing the number of recorded speckles, which can be accomplished either by decreasing their size or by opening the aperture of the camera, thus increasing the recorded area circle radius. The first option has a negative consequence if the speckle size is close to the detector pixel length. The expected contrast decreases abruptly when the speckle size falls below the sensor cell width, so that the latter poses a low bound for the former. 

Keeping $2c$ below $\frac 1 N \, \times$ (the recording area width) allows for the amplitudes of $N\,\times\,N$ points to be simultaneously measured. Therefore, the larger the radius $c$, the better the uncertainty, but the worse the horizontal resolution. 

Finally, the measuring range written in \Equa{mr} further restraints the possible values for $s$.

The mutually limiting relations between uncertainty, horizontal resolution and measuring range described in the previous lines are all reflected in the analysis, equations and figures of this paper.

\comentario{Speckle interferometric or holographic vibration measuring systems have a slightly better accuracy but need more restricted conditions to work and lack the flexibility and adaptability of the one decribed in this paper. 
}


\section{Mathematical results}\label{sec:mathresults}

This section lays out three mathematical results that are repeatedly referred to in the previous text. The first two of them have been worked specifically for this paper and the last one is a well established result of stochastic systems  theory. 
\begin{result}\label{ellipticK}
Let $H(x)$ be the autoconvolution of $h_1(x)$ defined at \Equa{h1def}, or, equivalently
\begin{equation}\label{defH}
H(t)=\frac{1}{\pi^2}\int_{-\zeta+\valorabst}^{\zeta}\frac{d x}{\sqrt{
(\zeta-x)(\zeta+x)(\zeta-\valorabst+x)(\zeta+\valorabst-x)
}}
\end{equation}
for $|t|\leq 2\zeta$ and $0$ otherwise. Then
\begin{equation}\label{newHL}
H(t)=\frac{4}{\pi^2(2\zeta+{\valorabst})}K(k) 
\end{equation}
where $K(k)$ is the complete elliptic integral of the first kind and 
\begin{equation}\label{paramK}
k=\frac{2\zeta-{\valorabst}}{2\zeta+{\valorabst}}
\end{equation}
for $|t|\leq 2\zeta$ and $0$ otherwise.
\end{result}

The result follows from a change of the integration variable to $\xi$ given by
\begin{equation}\label{chageintegr}
\xi =\frac{x-\frac{\valorabst}{2}}{\zeta-\frac \valorabst 2}
\end{equation}
the function $H(t)$ is thus expressed by
\bwt\begin{equation}\label{newH}
H(t)=\frac{1}{\pi^2(\zeta+\frac{\valorabst}{2})}\int_{-1}^1 \frac{d \xi}{\sqrt{(1-\xi^2)(1-k^2\xi^2)}}=\frac{4}{\pi^2(2\zeta+{\valorabst})}\int_{0}^1 \frac{d \xi}{\sqrt{(1-\xi^2)(1-k^2\xi^2)}}
\end{equation}\ewt
from which \Equa{newHL} is read straightforward.

\begin{result}\label{r1}
Let $J$ be the integral
\begin{equation}\label{az11}
J=\frac{1}{\pi^2 c^4}\intc F(x-x',y-y') \pupilf(c,x,y) \pupilf(c,x',y')d x d y d x' d y'
\end{equation}
where the circular pupil function $\pupilf$ is determined by $\pupilf(c,x,y)=1$ if $x^2 + y^2 \leq c^2$ and $0$ otherwise.

Let a function $L(x,y)$ be defined by
\begin{equation}\label{overl}
L(x,y)= \frac{2 }{\pi}\left(
\mbox{acos}\,\frac{\sqrt{x^2+y^2}}{2c} - \frac{\sqrt{x^2+y^2}}{2 c}\sqrt{1-\displaystyle \frac{x^2+ y^2}{4 c^2}}
\right)
\end{equation}
for $x^2 + y^2 \leq 4 c^2$ and $0$ otherwise. 

Then
\begin{equation}\label{az12}
J=\frac{1}{\pi c^2}\iint F(\xi,\eta)L(\xi,\eta)d \xi d \eta
\end{equation}
\end{result}
This results follows from a change of variables
\begin{equation}\label{az13}
\left\{
\begin{array}{rcl}
\xi&=& x-x'\\
\eta&=& y-y'\\
\xi'&=& x+x'\\
\eta'&=& y+y'
\end{array}
\right.
\end{equation}
which defines a Jacobian
\begin{equation}\label{az14}
\frac{D(\xi,\eta,\xi',\eta')}{D(x,y,x',y')}=4
\end{equation}
so that 
\bwt\begin{equation}\label{az15}
J=\frac{1}{4 \pi^2 c^4}\iint F(\xi,\eta) \left( \iint \pupilf(c,\frac{\xi'+\xi}{2},\frac{\eta'+\eta}{2}) \pupilf(c,\frac{\xi'-\xi}{2},\frac{\eta'-\eta}{2}) d \xi' d \eta' \right) d \xi d \eta
\end{equation}\ewt
where the inner integral is the overlapping area of two circles in the $\xi',\eta'$ plane with centers at $(\xi,\eta),(-\xi,-\eta)$ and the same radius $2c$. Substitution for this area
\begin{equation}\label{az16}
A=8c^2\left(
\mbox{acos}\,\frac{\sqrt{\xi^2 + \eta^2}}{2c} - \frac{\sqrt{\xi^2 + \eta^2}}{2 c}\sqrt{1-\displaystyle \frac{\xi^2 + \eta^2}{4 c^2}}
\right)
\end{equation}
yields the expression searched.
\begin{result}\label{r2}
Let $i_0(x)$ be a Wide Sense Stationary (WSS) stochastic process  and $i(x)=i_0(x) \otimes H(x)$ be the convolution of $i_0$ with the function $H$
then
\begin{equation}\label{az17}
R(\xi)=H(\xi)\otimes H(-\xi)\otimes R_0(\xi)
\end{equation}
where
$
R(\xi)=\ev{i(x)i(x+\xi)}, R_0(\xi)=\ev{i_0(x)i_0(x+\xi)}
$
are the autocorrelation functions of $i(x),i_0(x)$, respectively.
\end{result}
This result is not original and can be found at \cite{papoulis-1} among others.



\end{document}